# Polyurethane-Inspired $CO_2$ Chemisorbent: Ab Initio Reaction Profiles

Vitaly V. Chaban

Independent Scientist, Russian Federation. Email address: vvchaban@gmail.com

**Abstract:**

Polyurethane (PU) and its numerous fine-tuned derivatives are widely employed as $CO_2$ scavengers thanks to (1) physisorption and (2) functionalization of the PU backbone with other $CO_2$ sorbents. In the present work, it has been unraveled why PU cannot exhibit $CO_2$ chemisorption, despite possessing the nitrogen docking sites and exhibiting strong electrostatic sorbent-sorbate interactions. Furthermore, a few types of spatial separation of the active sorption sites have been proposed to unleash the chemisorption functionality of PU. By comparing various structural modifications of PU by using the in-silico methodology, we have identified that $CO_2$ chemisorption by PU takes place in the case of implementing methyl and ethyl fragments between the oxygen and nitrogen atoms of PU. Herewith, the introduction of the ethyl moiety even makes $CO_2$ chemisorption energetically favorable relative to physisorption. The reported specific progress on materials design represents an obvious practical value for chemical engineers developing inexpensive $CO_2$ scavengers.

**TOC Graphic**

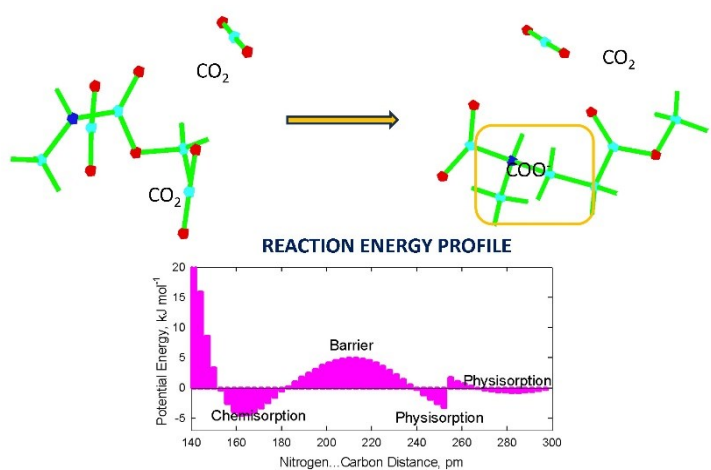

Insertion of the ethylene chain between the -N(H)- and -C(O)O- groups unleashes $CO_2$ chemisorption.

**Research Highlights**

The polyurethane chemisorption functionality has been unleashed.

The spatial separation of O and N atoms in PU enables $CO_2$ chemisorption.

The ethylene separator has been shown to offer the best chemisorption performance.

**Introduction**

The excess concentration of $CO_2$ in the atmosphere of the Earth represents a widely publicly discussed environmental peril nowadays due to its connection to global warming and climate alteration.[1-3] The consensus is that the emissions of human-associated greenhouse gases (methane and carbon dioxide) must be diminished to support a safer future. Numerous research teams around the globe develop novel materials exhibiting the functionality of $CO_2$ scavengers. The goals are to get an energetically efficient, cheap, and renewable $CO_2$ adsorbent and/or absorbent.[4]

Chemistry-wise, the problem of $CO_2$ capturing is challenging because of the thermodynamic and kinetic stabilities of this gas.[5-6] The enthalpic gain obtained due to covalent bonding with $CO_2$ must be large enough to compensate for the entropic penalty due to decreased conformational flexibility of the captured gas molecule. There are a relatively small number of covalent bonds, which $CO_2$ forms with its potential sorbents. Therefore, modern optimization of $CO_2$ scavengers relies on minor optimization of the scavenging material to preserve kinetic stability and minimize sorption and desorption costs at given thermodynamic conditions. Many natural and cheaply produced materials compete for being selected as $CO_2$ scavengers in industrial settings.[2]

Polyurethanes (PUs) designate a class of versatile polymers that are extensively used thanks to their noteworthy mechanical properties, durability, and resistance to environmental factors. PUs are synthesized by the reaction of polyols and isocyanates.[7] PUs are known to maintain different forms, including rigid foams, flexible foams, elastomers, adhesives, coatings, and sealants. The practical applications of PUs range from furniture and automotive parts to insulation materials and kitchenware as sponges.

A peculiar urethane functional group, -NHC(O)O-, combines three highly active chemical elements of the second period and the hydrogen atom.[8-11] The urethane group is formed by the reaction between an isocyanate, -NCO, and a hydroxyl group, -OH. The presence of the isocyanate moiety in the backbone is the defining feature of polyurethanes. The urethane group readily participates in hydrogen bonding, which significantly enhances the mechanical strength, elasticity, and thermal stability of the material. Urethane groups are responsible for the excellent adhesion properties of coatings and adhesives produced out of PU. The urethane group interacts with water, acids, and bases and can undergo hydrolysis under certain conditions.[12-13] Therefore, the chemical and physical properties of PUs degrade over time due to prolonged exposure to $CO_2$, moisture, and aggressive agents, which are frequently present in $CO_2$-containing streams. This could limit the long-term effectiveness of PUs in $CO_2$ scavenging applications.

The interaction with $CO_2$ takes place through electrostatic attraction forces.[14] This leads to physical adsorption, whose intensity depends on the surrounding chemical structures possessing nucleophilic and electrophilic reaction sites. PUs can be functionalized to go beyond the chemical functionality of the backbone. For instance, amine moieties or anionic fragments derived from ionic liquids can be grafted to enhance the affinity of the material to $CO_2$. Carbamate and bicarbonate species emerge upon $CO_2$ chemisorption while utilizing such fine-tuned derivatives.[14]

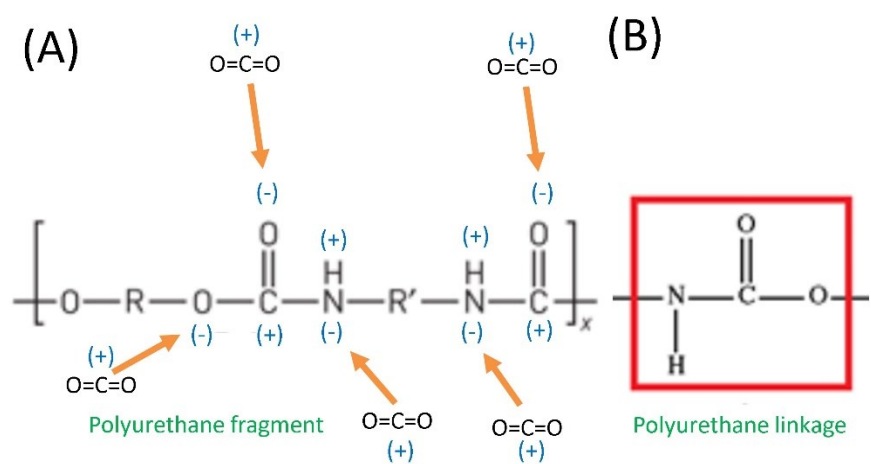

Figure 1. (A) Expected physisorption interactions of PU fragment with the $CO_2$ molecules. (B) Polyurethane linkage, -HNC(O)O-.

Certain types of open-cell PU foams with high surface areas and hierarchical porosity offer a great promise to physically adsorb $CO_2$ through the combining effect of Coulombic forces and van der Waals forces.[15-17] The large surface area provided by the PU foam structure fosters contact between the $CO_2$ molecules and irregular polymer. Non-covalent interactions with a few PU chains simultaneously boost the efficiency of the capture. The $CO_2$ capture via physisorption is less selective but can still be effective for capturing $CO_2$ under specific conditions. Physisorption is widely perceived as a method with lower desorption energy costs as compared to $CO_2$ chemisorption. Tailorability is the cornerstone advantage of PUs as a functional material. Compared to sorptive materials for competitive $CO_2$ capture technologies, PUs are produced at a relatively low cost. Cheapness makes PUs one of the most attractive options for large-scale deployment.

In the present work, we use a computational methodology to solve two paramount issues related to the $CO_2$ chemisorption of PUs. First, we prove the principal impossibility of the non-functionalized PU to chemically absorb $CO_2$ due to the repulsive interaction between the oxygen atom of $CO_2$ and the oxygen atom of the carboxyl moiety at some point along the path of $CO_2$ towards the imine moiety of PU. We describe the unfavorable intermolecular O…O interaction and precisely characterize it in energetical and geometrical units. Second, we investigate minor structural modifications to potentially unleash the $CO_2$ chemisorption potential of PUs. Such straightforward modifications were determined at the level of in-silico atomistically precise design of materials and comprehensively characterized in terms of reaction thermochemistry, polarities of the reacting sites, and structural descriptors.

**Methodology and Methods**

The reaction energy profiles, containing the transition state, were obtained by forcibly moving the assigned reaction coordinate from the $CO_2$ gaseous state through the state of $CO_2$ physisorption toward the state of $CO_2$ chemisorption. The most intuitive reaction coordinate corresponding to the $CO_2$ chemisorption, carbon($CO_2$)…nitrogen(PU) distance, was selected. The vibrational frequency calculations and the intrinsic reaction coordinate simulations were afterward employed to confirm that the transition state units the physisorption state to the chemisorption state. A valid transition state must represent the first-order saddle point and contain a single vibrational frequency in its vibrational profile.

The reaction coordinate was propagated forcibly by decreasing the value of the linear reaction coordinate stepwise, by 3 pm, in all simulated chemical compositions. At each point of the energy profile, the value of the reaction coordinate was kept rigidly fixed. In turn, all other degrees of freedom of the system were allowed to adapt to the running reaction coordinate. This procedure is well-known as a relaxed energy scan in various contexts.

All recorded reaction profiles contain at least, one stationary point. This is the state of $CO_2$ physisorption. If an additional minimum state exists at the reaction coordinate value smaller than 200 pm, it is assigned as a $CO_2$ chemisorption state. The mentioned distance criterion is computed as 1.4 multiplied by the expected length of the carbon($CO_2$)-nitrogen(PU) covalent bond. Furthermore, this criterion suggests an expected location of the reaction transition state This is a rule of thumb that the bond formation/bond breaking transition state is located around 130-140% of the equilibrium bond length.

The stationary points are characterized by covalent bond lengths, non-covalent interatomic distances, covalent angles, and partial atomic electrostatic charges as measures of reaction sites' nucleophilicity and electrophilicity. The partial atomic electrostatic charges are derived by fitting the electrostatic potential for electrons and atomic nuclei using the Merz-Kollman algorithm.[18] The thermodynamic potentials – Gibbs free energy, enthalpy, and configurational entropy[19] are computed at various temperatures and pressures – for $CO_2$ physisorption and $CO_2$ chemisorption using the sequence of Hess' law. The electronic energy-based heights of the obtained activation barriers are corrected correspondingly. Herewith, the physisorption thermodynamics is computed versus an ideal gas state of $CO_2$, i.e. the $CO_2$-$CO_2$ attraction is ignored. The corrections for zero-point energy are computed in every stationary point through the harmonic frequency analysis. The anharmonic frequency analysis is also performed in several systems to assess the cost of the approximation for the accuracy of the results. The energetic effects of the electronic excitations are not included in these calculations.[19]

Unrestricted hybrid density functional theory (HDFT), as implemented in the range-separated exchange-correlation meta-GGA functional M11,[20] is used. The molecular wave functions rely on the primitive orbitals belonging to the atom-centered triple-zeta split-valence polarized Def2-TZVP basis set.[21] The dispersive attraction is included inherently.[20] The convergence criterion within the self-consistent field procedure was set to $10^{-7}$ hartree. The rational function optimization algorithm is used to optimize partially restrained geometries along the proposed minimum reaction path. The following geometry convergence criteria were chosen: 120 kJ/nm for the largest force and 40 kJ/nm for root-mean-squared force.

The HDFT calculations were conducted in GAMESS.2020.[22] The in-house scripts were utilized to carry out all other analysis routines. The Avogadro drag-and-drop software was used to draw preliminary molecular geometries of the simulated chemical structures.[23]

**Results and Discussion**

In this section, we first show that pristine PU does not chemically absorb $CO_2$. The scavenger's performance of PU is exclusively due to numerous pairwise electrostatic attractions, primarily involving electrophilic carbon atoms in $CO_2$ and carboxylate moieties. Next, we modify the structure of PU aiming to avoid unfavorable intermolecular interactions that prevent chemical absorption. The modifications appear to be successful and unleash the desired new functionality, which is characterized in terms of kinetic and thermochemical descriptors.

**Chemisorption of $CO_2$ in Polyurethane**

Figure 2 depicts the absence of $CO_2$ chemisorption in pristine PU and visualizes key non-covalent interactions, which influence this sort of system behavior. No activation energy barrier that could separate the state of $CO_2$ chemisorption from the state of $CO_2$ physisorption was identified. Therefore, no chemisorption state exists as well. Even if such a state is attained during thermal motion at the temperature of $CO_2$ capture, it is spontaneously destroyed during microscopic time frames at any finite translational temperature.

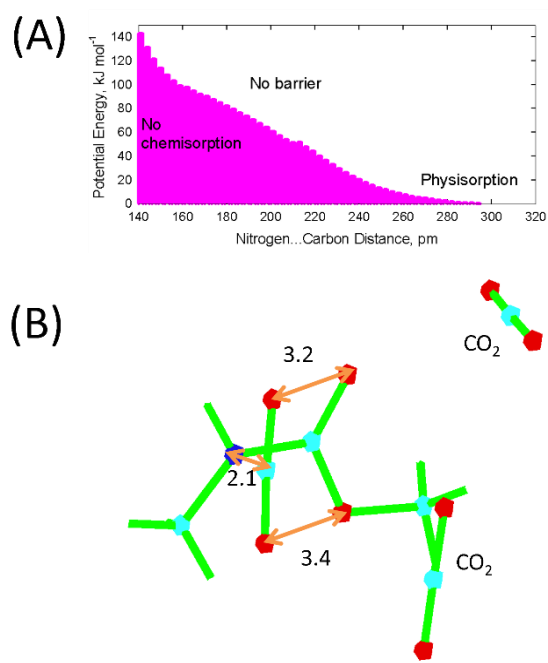

Figure 2. (A) Computed $CO_2$ chemisorption energy profile for PU. (B) Electrostatic oxygen-oxygen repulsions, preventing $CO_2$ chemisorption by PU. The in-plot distances are given in Ångstroms.

The partially optimized geometry of the simulated system, containing one urethane unit and three $CO_2$ molecules, is depicted in Figure 2b to rationalize the reaction energy profile in Figure 1a. At the reaction coordinate of 214 pm, the two oxygen atoms of the approaching $CO_2$ meet the two oxygen atoms of the urethane linkage, which is an inherent part of the sorbent. The corresponding distances amount to 3.2 and 3.4 Å. Since all four oxygen atoms mentioned above are strongly nucleophilic, they repel strongly at relatively small interatomic distances. Their repulsions directly influence the shape of the reaction energy profile and suppress the chemisorption minimum. A logical conclusion would be that structural modification of the sorbent can unleash the currently inaccessible reaction.

Two other $CO_2$ molecules included in the same system cannot be captured chemically. They were captured electrostatically by the carboxylate moiety of PU. In both cases, the oxygen atoms of PU interact with the carbon atoms of two $CO_2$ molecules. Both equilibrium distances C…O equal 2.9 Å, which corresponds to the electrostatic attraction of medium strength. Table 1 summarizes thermodynamic potentials corresponding to $CO_2$ physisorption.

Table 1. The thermodynamic potentials of $CO_2$ physisorption by methylene-modified PU. The normalization of energies throughout this work is provided per one mole of $CO_2$.

| Temperature, K | Pressure, bar | dG, kJ/mol | dH, kJ/mol | -TdS, kJ/mol | dS, J/mol/K |
|---|---|---|---|---|---|
| 280 | standard | 23.50 | -7.95 | 31.45 | -112.3 |
| 290 | standard | 24.62 | -7.83 | 32.45 | -111.9 |
| standard | standard | 25.53 | -7.74 | 33.27 | -111.6 |
| 310 | standard | 26.85 | -7.60 | 34.45 | -111.1 |
| 320 | standard | 27.96 | -7.49 | 35.44 | -110.8 |
| 330 | standard | 29.06 | -7.36 | 36.43 | -110.4 |
| standard | 10 | 19.82 | -7.74 | 27.56 | -92.45 |
| standard | 50 | 15.83 | -7.74 | 23.57 | -79.05 |
| standard | 100 | 14.11 | -7.74 | 21.85 | -73.30 |
| standard | 200 | 12.39 | -7.74 | 20.13 | -67.52 |
| standard | 500 | 10.12 | -7.74 | 17.86 | -59.91 |

The physisorption is enthalpically favored and entropically forbidden over the entire temperature and pressure ranges. Elevated temperature decreases $CO_2$ sorption. In turn, elevated pressure fosters physisorption. The most favorable thermodynamics, in terms of Gibbs free energy, is seen at 298.15 K and 500 bar. When a more realistic pressure is applied, in a practical context, dG suffers insignificantly, with 12 kJ/mol under 200 bar and 14 kJ/mol under 100 bar. Assuming a porous structure of PU that can accommodate multiple captured gas molecules, even more favorable thermodynamics can be expected. Indeed, PU scavengers exhibit rather competitive $CO_2$ capacities in recent experiments.

**Methylene and Ethylene Modifications of Polyurethane**

Since chemisorption is impossible in PU, we modified the sorbent structure by separating oxygen and nitrogen atoms from one another using the methylene linkage and the ethylene linkage. In this way, we eliminate thermodynamically unfavorable O…O electrostatic repulsions. Figure 3 compares the physisorption by PU and two modified chemical structures.

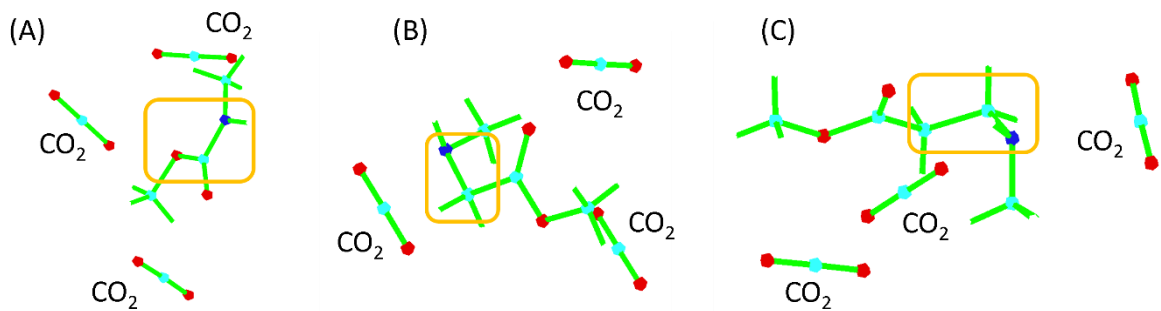

Figure 3. The optimized geometries depicting physisorption and chemisorption of $CO_2$ by (A) pristine PU unit; (B) modified PU by adding a methylene separator; (C) modified PU by adding an ethylene separator. The urethane linkage and its modifications are highlighted by orange rectangles.

The $CO_2$ molecules are coordinated by one nitrogen and two oxygen atoms. In all cases, nucleophilic centers of PU and modified PU are responsible for the observed $CO_2$ capture. We found that chemisorption through carboxamidation was not spontaneous. Otherwise, it would take place directly during the simulated geometry optimization following the rational-function-optimization algorithm employed. Figure 4 depicts the chemisorption of $CO_2$ by the methylene-modified scavenger and exemplifies the geometry of the chemisorbed $CO_2$ molecule at the nitrogen site.

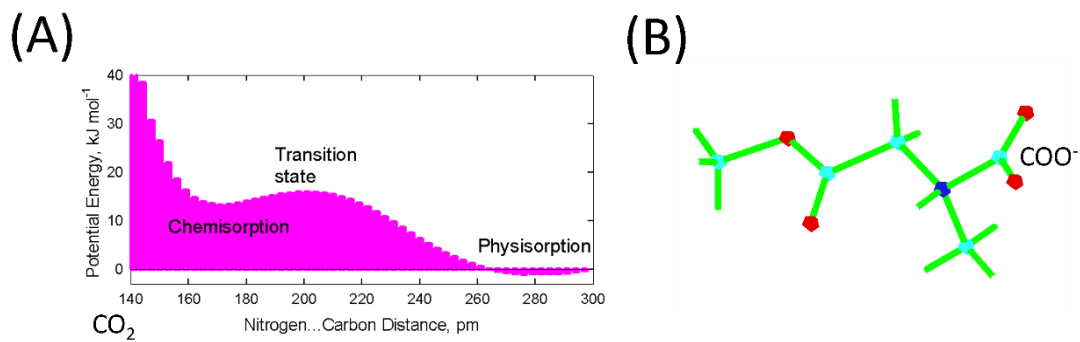

Figure 4. (A) Reaction energy profile for $CO_2$ chemisorption by methylene-modified structure. (B) The chemisorbed $CO_2$ molecule via a new PU-inspired scavenger.

The transition state is located at the reaction coordinate of 201 pm, which is C…N distance. The height of the barrier is 17 kJ/mol in terms of electronic energy. The height of an activation barrier is evaluated versus the lowest-energy state located on the right. In our case, it is a state of the $CO_2$ physisorption. The imaginary frequency of the saddle point amounts to i151 cm$^{-1}$.

The chemisorbed $CO_2$ state was found at the reaction coordinate of 171 pm. The energy of the simulated system relative to the physisorption energy was 14 kJ/mol. Such small involved energies suggest that $CO_2$ chemisorption is readily accessible even at standard conditions. In the meantime, chemisorption is enthalpically forbidden relative to physisorption. It should represent a reversible reaction that can be acceptable in the context of less expensive sorbent regeneration.

Thermochemistry of $CO_2$ chemisorption (Table 2) suggests that pressure increase does not boost chemisorption. Furthermore, chemisorption is entropically forbidden, although to a lesser extent than physisorption. The enthalpic gain of the carboxamidation reaction is within 1 kJ/mol versus that of $CO_2$ physisorption. Therefore, the chemisorption functionality was activated by a simple structural modification of the sorbent unit. A larger separating moiety can provide more efficient charge separation and enhance the chemisorption of $CO_2$.

Table 2. Thermochemical potentials of $CO_2$ chemisorption by methylene-modified PU.

| Temperature, K | Pressure, bar | dG, kJ/mol | dH, kJ/mol | -TdS, kJ/mol | dS, J/mol/K |
|---|---|---|---|---|---|
| 280 | standard | 8.55 | 0.43 | 8.12 | -28.99 |
| 290 | standard | 8.83 | 0.37 | 8.47 | -29.20 |
| standard | standard | 9.07 | 0.32 | 8.76 | -29.37 |
| 310 | standard | 9.42 | 0.25 | 9.17 | -29.57 |
| 320 | standard | 9.72 | 0.20 | 9.52 | -29.76 |
| 330 | standard | 10.02 | 0.14 | 9.87 | -29.92 |
| standard | 10 | 9.07 | 0.32 | 8.76 | -29.37 |
| standard | 50 | 9.07 | 0.32 | 8.76 | -29.37 |
| standard | 100 | 9.07 | 0.32 | 8.76 | -29.38 |
| standard | 200 | 9.08 | 0.32 | 8.76 | -29.38 |
| standard | 500 | 9.07 | 0.32 | 8.76 | -29.37 |

Table 3 investigates the thermodynamics of $CO_2$ physisorption by using an ethylene-modified PU linkage. Indeed, a slight dG decrease took place. Specifically, each dG became less positive by roughly 2 kJ/mol. Although the exemplified energy gains are marginal, they signify a positive effect of spatially separating nucleophilic interaction sites from one another. Also, the effects originating from many urethane units may sum up into a substantial number for an entire PU chain.

Table 3. The thermodynamic potentials of $CO_2$ physisorption by ethylene-modified PU.

| Temperature, K | Pressure, bar | dG, kJ/mol | dH, kJ/mol | -TdS, kJ/mol | dS, J/mol/K |
|---|---|---|---|---|---|
| 280 | standard | 21.58 | -9.84 | 31.41 | -112.19 |
| 290 | standard | 22.70 | -9.72 | 32.42 | -111.79 |
| standard | standard | 23.61 | -9.63 | 33.23 | -111.47 |
| 310 | standard | 24.93 | -9.49 | 34.42 | -111.03 |
| 320 | standard | 26.04 | -9.38 | 35.41 | -110.66 |
| 330 | standard | 27.14 | -9.26 | 36.40 | -110.31 |
| standard | 10 | 17.90 | -9.63 | 27.53 | -92.32 |
| standard | 50 | 13.91 | -9.63 | 23.53 | -78.94 |
| standard | 100 | 12.19 | -9.63 | 21.82 | -73.18 |
| standard | 200 | 10.48 | -9.63 | 20.10 | -67.42 |
| standard | 500 | 8.20 | -9.63 | 17.83 | -59.79 |

Ethylene-modified PU is more successful in $CO_2$ capture (Figure 5). For instance, the chemisorption barrier decreases thrice as compared to the methylene-modified scavenger. The state of chemisorption is more expressed versus the transition state. Two physisorption states emerge, which are linked to a more crowded molecular environment in the proximity of the nitrogen reaction site. The ethylene separator blocks strong electrostatic interactions within the PU linkage. As a result, the thermodynamic potentials favor the physisorption and chemisorption of $CO_2$ to a larger extent.

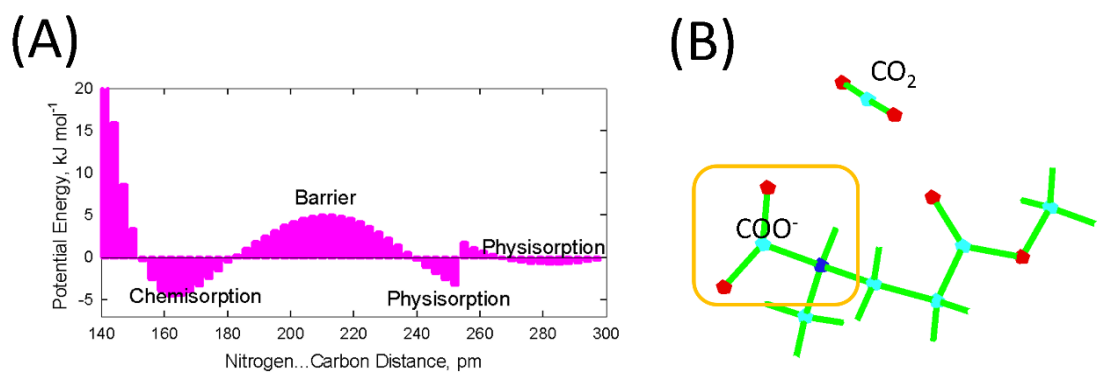

Figure 5. (A) Reaction energy profile for $CO_2$ chemisorption by ethylene-modified structure. (B) The chemisorbed $CO_2$ molecule and physisorbed $CO_2$ molecule via a new PU-inspired scavenger.

The imaginary frequency characterizing the transition state in the ethylene-modified PU equals 187 cm$^{-1}$. This is reasonably similar to the transition state in the methylene-modified PU reported in this work above. The location of the activation barrier is 210 pm. In turn, the chemisorbed $CO_2$ molecules can be observed at the C…N distance of 163 pm. This bond length is in excellent agreement with conventional C-N single covalent bond lengths in numerous organic compounds. Table 4 summarizes $CO_2$ chemisorption by the ethylene-modified novel scavenger.

Table 4. Thermochemical potentials of $CO_2$ chemisorption by ethylene-modified PU.

| Temperature, K | Pressure, bar | dG, kJ/mol | dH, kJ/mol | -TdS, kJ/mol | dS, J/mol/K |
|---|---|---|---|---|---|
| 280 | standard | 16.60 | 3.51 | 13.09 | -46.75 |
| 290 | standard | 17.07 | 3.33 | 13.74 | -47.39 |
| standard | standard | 17.46 | 3.18 | 14.28 | -47.90 |
| 310 | standard | 18.03 | 2.97 | 15.06 | -48.59 |
| 320 | standard | 18.52 | 2.79 | 15.72 | -49.14 |
| 330 | standard | 19.01 | 2.62 | 16.39 | -49.68 |

The $CO_2$ chemisorption capability, unlocked in the present work, transforms the urethane linkage into an anion. Deprotonated carbamate moiety emerges. The excess electron is localized on the oxygen atoms of carboxylate. Therefore, the stability of the system and, hence, the thermochemistry of the chemisorption reaction can be boosted in acidic media. In the reported simulations, we did not immediately include protons or counterions. Even without this sort of stabilization, the stationary points corresponding to the chemisorbed $CO_2$ molecules exist in both modified species.

Physisorption plays a substantial role in $CO_2$ capture by the modified PU scavengers, as well as by pristine PU. We hereby found three stationary points for ethylene-modified PU and $CO_2$ molecules (Figure 6). The stationary point, at which $CO_2$ coordinates the nitrogen atom of PU (C…N distance of 2.8 Å), can readily transition to the chemisorption stationary point (covalent bond length of 1.6 Å) at room conditions, according to the computed energy barriers discussed above. In turn, the states of $CO_2$ physisorption correspond to the electrostatic attraction of the carbon atom of $CO_2$ to the oxygen atoms of the carboxylate moieties.

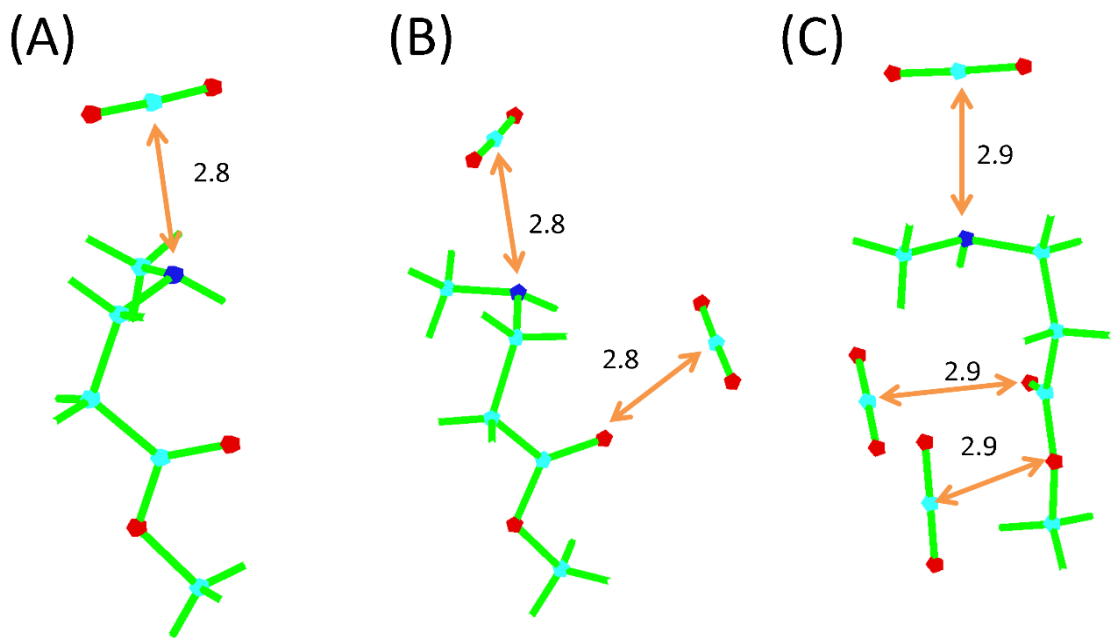

Figure 6. CO$_2$ physisorption geometries at ethylene-modified PU: (A) one CO$_2$ molecule; (B) two CO$_2$ molecules; (C) three CO$_2$ molecules. The in-plot distances are given in Ångstroms.

**Atomic Nucleophilicities and Electrophilicities along the Reaction Coordinate**

Figures 7-8 demonstrate how the partial atomic charges localized on the reacting centers evolve during CO$_2$ chemisorption. In the beginning, the carbon atom of CO$_2$ represents a typical electrophilic reaction center. Its partial charge equals ~0.7e. Whereas, the nitrogen atoms belonging to the methylene- and ethylene-modified PU species represent nucleophilic reaction centers.

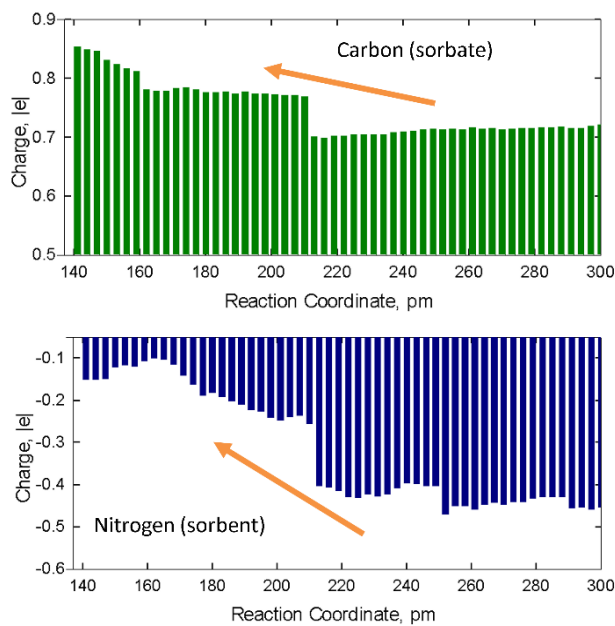

Figure 7. Partial atomic charges on the carbon and nitrogen reaction centers along the simulated $CO_2$ chemisorption reaction coordinate in pristine PU. The charges were computed by fitting an immediate electrostatic potential by a set of atomic coordinates.

Importantly, the case of ethylene separator turned out to be much more efficient. It increased the nucleophilicity of the nitrogen atom nearly twice compared to that in the case of the methylene separator. Compare -0.4e…-0.5e to almost -0.8e. This observation, made in terms of partial electrostatic charges, highlights that a very essential alteration in the electronic properties of PU-modified sorbent can be achieved via a tiny structural adjustment. It is also important to note that partial charges change according to different patterns along the reaction coordinates in methylene-modified and ethylene-modified sorbents.

As the reaction coordinate value approaches the state of chemisorption, the nitrogen atom of the ethylene-modified PU completely loses its initial nucleophilicity and even becomes somewhat electrophilic in the state of $CO_2$ chemisorption. In turn, the nitrogen atom upon methylene modification also substantially loses nucleophilicity but still retains a modest fraction of it in the state of carbamate moiety.

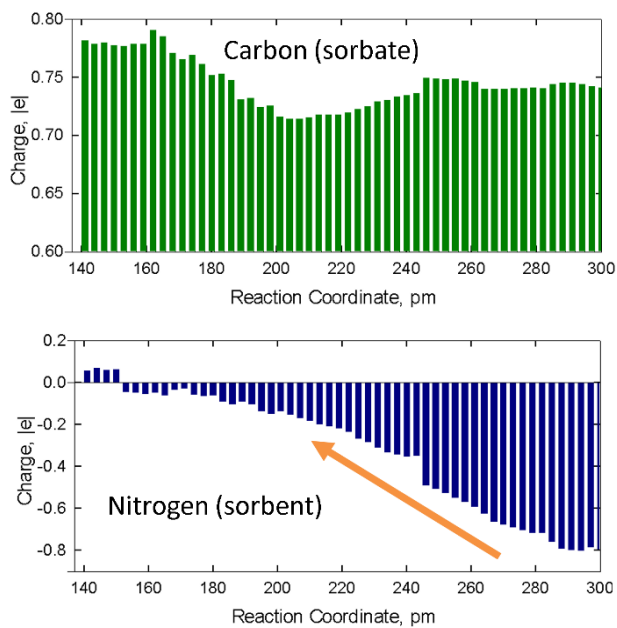

Figure 8. Partial atomic charges on the carbon and nitrogen reaction centers along the simulated $CO_2$ chemisorption reaction coordinate in ethylene-modified PU. The charges were computed by fitting an immediate electrostatic potential by a set of atomic coordinates.

The evolution of nucleophilicity and nucleophilicity of the directly chemically reacting atoms, nitrogen and carbon, upon their mutual approach, are in full agreement with the corresponding reaction energy profiles. These results suggest that an ethylene separator is most efficient in circumventing O…O undesirable repulsions on the way of carbamate synthesis.

**Chemisorption Activation Barriers versus Temperature**

The transition state represents the first-order saddle point, in which all forces are negligible. All descriptors characteristic of any stationary point apply to the transition state point. The activation energy barrier is a difference between the stational point energy and the closest minimum point energy. Tables 5-6 investigate how $CO_2$ chemisorption activation barriers depend on the temperature of the reaction. They also include zero-point energy corrections.

Table 5. Activation barriers of $CO_2$ chemisorption by methylene-modified PU. The effect of pressure is marginal in the usually experimentally used range of 1-200 bar.

| Temperature, K | Pressure, bar | dG, kJ/mol | dH, kJ/mol | -TdS, kJ/mol | dS, J/mol/K |
|---|---|---|---|---|---|
| 280 | standard | 24.68 | 15.33 | 9.35 | -33.41 |
| 290 | standard | 25.02 | 15.21 | 9.81 | -33.83 |
| standard | standard | 25.30 | 15.11 | 10.18 | -34.16 |
| 310 | standard | 25.71 | 14.97 | 10.74 | -34.63 |
| 320 | standard | 26.05 | 14.85 | 11.20 | -35.00 |
| 330 | standard | 26.40 | 14.73 | 11.67 | -35.36 |

The computed energy barriers are fairly small in all cases. It is obvious after the analysis of the thermodynamic tables that ethylene-modified PU outperforms methylene-modified PU by about 6 kJ/mol. Temperature increase insignificantly deteriorates the chemisorption reaction page through elevating barriers. In turn, the effect of pressure was found to be marginal (not shown). Mind, however, that pressure does influence $CO_2$ physisorption, as we discussed above.

Table 6. Activation barriers of $CO_2$ chemisorption by ethylene-modified PU. The effect of pressure is marginal in the usually experimentally used range of 1-200 bar.

| Temperature, K | Pressure, bar | dG, kJ/mol | dH, kJ/mol | -TdS, kJ/mol | dS, J/mol/K |
|---|---|---|---|---|---|
| 280 | standard | 18.61 | 10.31 | 8.30 | -29.64 |
| 290 | standard | 18.90 | 10.21 | 8.70 | -29.99 |
| standard | standard | 19.15 | 10.13 | 9.02 | -30.27 |
| 310 | standard | 19.51 | 10.01 | 9.50 | -30.64 |
| 320 | standard | 19.82 | 9.91 | 9.90 | -30.95 |
| 330 | standard | 20.13 | 9.82 | 10.31 | -31.24 |

**Conclusions and Final Considerations**

In the present work, a new $CO_2$ scavenger was elaborated by unlocking the chemisorption capability of PU. Specifically, thermodynamically unfavorable oxygen-oxygen repulsive interactions due to the rigidity of the sorbent's structure were identified by recording a minimum-energy reaction path and eliminated. Small separators, like methylene and ethylene groups, were inserted between the nitrogen and carbon atoms of the PU backbone. As a result, a new stationary point emerged in the simulated system corresponding to a chemically linked $CO_2$ molecule to the urethane linkage.

$CO_2$ chemisorption follows a well-known mechanism of carboxamidation. Whereas the added performance of the sorbents brings only 5 kJ/mol of energy to the system in the case of the ethylene bridge and even weakens the stability of the simulated system in the case of the methylene bridge, a novel functionality of the scavenger contributes to the $CO_2$ sorption flexibility and versatility. The physisorption of $CO_2$ by the new sorbents is slightly exothermic, whereas the chemisorption effect is very close to zero. All reactions are entropically forbidden which is usual for gas sorption processes.

The ab initio simulations reveal that three $CO_2$ molecules can be fixed by one unit of each of the new scavengers. Herewith, two gas molecules are physisorbed thanks to O…C electrostatic coupling, and one gas molecule is chemisorbed thanks to the covalent N-C bond. $CO_2$ physisorption is corroborated by its equilibrium distances from the sorbents, which are inferior to 3 Å. The drawback of the new materials is the inability to directly synthesize them out of PU. Indeed, methylene and ethylene moieties cannot be directly inserted to separate oxygen and nitrogen from the urethane linkage. In the meantime, such separators represent thermodynamically stable structures and in no way undermine the stabilities of the exemplified scavengers as compared to PU. Synthetic efforts are hereby urged to verify the effects of the modified PU structures on scavengers' performance.

**Acknowledgments**

**Conflict of Interest**

The author hereby declares no financial interests and professional connections that might bias the interpretations of the obtained results.

**Data Availability Notice**

The data supporting this article have been included in the publication.

**Credit Author Statement**

V.V.C.: Conceptualization; Methodology Development; Validation; Formal analysis; Investigation; Resources; Data Curation; Writing - Original Draft; Writing - Review & Editing; Visualization Preparation; Supervision; Project administration.